\documentclass[12pt]{article}
\pdfoutput=1

\usepackage{color}
\usepackage{graphicx}
\usepackage{amsmath}
\usepackage{amssymb}
\usepackage{xspace}
\usepackage{slashed} %
\usepackage[small]{subfigure}
\usepackage[numbers,compress]{natbib}
\usepackage{epigraph}
\usepackage{tikzsymbols} %

 \usepackage{etoolbox}
\makeatletter
\newlength\epitextskip
\pretocmd{\@epitext}{\em}{}{}
\apptocmd{\@epitext}{\em}{}{}
\patchcmd{\epigraph}{\@epitext{#1}\\}{\@epitext{#1}\\[\epitextskip]}{}{}
\makeatother

\usepackage[hyperfootnotes=false]{hyperref}

\usepackage{mciteplus} 

\newlength{\abstractwidth}
\renewcommand{\title}[1]{\vbox{\center\bf{\Large{#1}}}\vspace{5mm}}
\renewcommand{\author}[1]{\vbox{\center#1}\vspace{5mm}}
\newcommand{\address}[1]{\vbox{\center\em#1}}
\newcommand{\email}[1]{\vbox{\center\tt#1}\vspace{5mm}}

\newcommand{\be}{\begin{equation}}
\newcommand{\ee}{\end{equation}}
\newcommand{\bi}{\begin{itemize}}
\newcommand{\ei}{\end{itemize}}

\renewcommand{\le}{\left}
\newcommand{\ri}{\right}

\begin{document}

\begin{titlepage}
\rightline{MIT-CTP/5269}
\begin{center}
\hfill \\
\hfill \\
\vskip .5cm

\title{Why is AI hard and Physics simple?}

\author{Daniel A. Roberts}

\address{
    Center for Theoretical Physics and Department of Physics,\\ Massachusetts
  Institute of Technology,\\ Cambridge, MA 02139, USA
 \\~\\
 Salesforce, Cambridge, MA 02139, USA
}

\email{drob@mit.edu}

\end{center}

\begin{abstract}

We discuss why AI is hard and why physics is simple. %
We discuss how physical intuition and the approach of theoretical physics can be brought to bear on the field of artificial intelligence and specifically machine learning. We suggest that the underlying project of machine learning and the underlying project of physics are strongly coupled through the principle of sparsity, and 
we call upon theoretical physicists to work on AI \emph{as physicists}.
As a first step in that direction, we discuss an upcoming book on the principles of deep learning theory that attempts to realize this approach \cite{Principles}.

\end{abstract}

\end{titlepage}

\tableofcontents

\section{Some Questions}\label{sec:introduction}
\epigraph{[T]he mathematical formulation of the physicist's often crude experience leads in an uncanny number of cases to an amazingly accurate description of a large class of phenomena.}{Eugene Wigner, ``The Unreasonable Effectiveness of Mathematics in the Natural Sciences''}

\noindent{}In a famous essay \cite{wigner1960unreasonable}, Eugene Wigner observed the important role that mathematical structure plays in describing physical theories. Wigner was both a mathematician and a physicist, and most of the engagement with his essay
focuses on the ``mathematical formulation'' part of the quote, specifically on the puzzle of why mathematics and physics should have anything to do with each other.

Our engagement will be different. Ignoring the mathematician's appeal to rigorous (re)formulation, we focus on the latter part of the quote: ``The physicist's often crude experience leads in an uncanny number of cases to an amazingly accurate description of a large class of phenomena.''

Why is this the case? What can physicists bring to other fields that have a natural description in terms of mathematics, such as artificial intelligence? %
Can they bring anything that isn't already known by mathematicians?
Given the accumulated success of the mathematical formulation of physics, perhaps we might hope to also find 
physics in other useful mathematical structures?

In this essay, we will attempt to identify the aspects of artificial intelligence (AI) that make it so hard (\S\ref{sec:ai-hard}). 
In particular, our focus will be on machine learning approaches to artificial intelligence in which functions are \emph{trained} rather than explicitly \emph{programmed}.
We will find that the difficulty is in precisely understanding the sparse mathematical structure 
that must be imposed upon any prospective 
machine learning model in order to make it tractable. This structure can be somewhat understood through a combination of intuition and analogy to human intelligence, but so far has appeared to resist any principled theoretical approach.%

Setting that aside for a moment, we will then comment on why a ``physicist's crude experience'' leads to such accurate %
and predictive theories (\S\ref{sec:physics-simple}). The physicist's engagement with math is not that of the mathematician, because the physicist is constrained by both theory \emph{and} experiment. It is precisely the interplay between the two that gives the physicist's  mathematical formulations such power in their descriptive abilities. In short, we will explain 
why physics is simple enough to enable immense progress in understanding
phenomena in nature.\footnote{
    Please note that the notion of \emph{simplicity} to which we are appealing is not at all meant to suggest that physics is \emph{trivial}. Instead, we mean it as a compliment: we have the utmost respect for the work of physicists. This essay is an apologia for physicists doing machine learning \emph{qua physicists}; it is meant to interrogate what it is about the approach or perspective of physicists that allows them to reach so far in explaining fundamental phenomena and then consider whether that same approach could be applied more broadly to understanding intelligence.
}

In many ways, machine learning and physics are two sides of the same coin.
The program of machine learning is the selection and fitting of mathematical models to describe some desired phenomena usually encoded in a probability distribution.  Traditionally, that is also a description of the program of physics, provided we specify that the phenomena are \emph{natural}.  

Perhaps with foresight, Wigner speaks of the physicist's success in describing phenomena, without bothering to limit that class to those that are natural.\footnote{
    One could argue that the field of condensed matter physics is mainly engaged in studying \emph{unnatural} phenomena. And thermodynamics grew out of the investigation of the \emph{artificial machines} of the steam age, eventually finding an underlying microscopic description in statistical mechanics. 
    All together, this is rather suggestive that the language and methods of physics might be particularly well suited for learning principles of machine learning. 
    We thank Sho Yaida for making and developing this point about physics and unnatural phenomena.
}  
It's our thesis that the tools and intuitions of the physicist can have a much broader domain of applicability, and 
we hope that more theoretical physicists will be open to making contributions to the field of AI by studying machine learning models \emph{as physicists}. Given the importance of tying such theoretical work directly to experiment, we anticipate that such efforts could have a direct impact on models at the forefront of AI research.

In light of our discussion of simplicity in physics, 
in \S\ref{sec:do} we 
speculate on the ways in which 
physics 
approaches to understanding
intelligence 
might be useful, and  in \S\ref{sec:ETofDL} we connect our ideas to the approach of an upcoming book \cite{Principles} that attempts to %
develop some principles of 
deep learning 
theory 
as a concrete  
example of what such a contribution could look like.

\subsection{Why is AI hard?}\label{sec:ai-hard}

\noindent{}Artificial Intelligence is hard because there is no such thing as a free lunch. 

This is the well-known oft-quoted statement 
that, when you average over all possible machine-learning problems, 
all learning algorithms are equivalent \cite{wolpert1996lack,wolpert1997no}.

For example, the number of possible $n$-pixel images, with each pixel being either black or white, is $2^n$: an exponentially large number. If we give each image one of two labels, then the number of possible ways of labeling the set of images is $2^{2^n}$: a doubly-exponentially large number. To get a sense of what this means, for even modest values of $n$ this is a absurdly large number: since $2^{2^{9}} \sim 10^{154}$, there are a preposterously greater number of ways to classify the set of 9-bit black and white images than there are atoms in the observable universe $\sim 10^{80}$ \cite{eddington2012philosophy}. 

With that in mind, if each image could in principle receive a different label without leveraging any notion of structure or relationships between the images -- i.e.~if the labels don't correlate in any way with the detailed properties or \emph{features} of the image -- then the best one could hope to do is memorize each image with its label. 
This is the essence of the no-free-lunch theorem: \emph{(i)} 
the task of memorizing images with their labels grows so absurdly quickly that it's more or less impossible for any interesting problem size, and \emph{(ii)} it is impossible to devise a scheme or algorithm that does better than such explicit memorization.

This seems pretty bleak. One reading of this theorem seems to be that no matter how we try to improve our methods and tools for machine learning, we can never do better than random. A learning algorithm that predicts the label $y$ on a previously unseen image $x$ and another algorithm that predicts the opposite label $\neg y$ will do equally as well in the long run if 
we average over all possible unseen data points $(x,y)$. 
Thus, we might as well just make a random prediction; in other words, pure and complete \emph{generalization} is impossible.

Yet humans learn very efficiently \cite{gopnik1999scientist,griffiths2006optimal,vul2014one}, and machine learning -- in particular realized via  artificial-neural-network-based function approximation: \emph{deep learning} --  appears to work as well \cite{lecun2015deep,ImageNet2012,goodfellow2014generative,mikolov2013efficient,mikolov2013distributed,BERT2018,radford2019language,Brown2020LanguageMA,mnih2015human,silver2016mastering,silver2018general,berner2019dota,starcraft,muzero,alphafold2prelim}. Once you 
return from absorbing the content in both citation dumps, you have to admit that there's a bit of tension between the no-free-lunch theorem and the apparent success of intelligence, both human and artificial.

In fact, the no-free-lunch theorem does not mean that there is no point in trying; rather, it implies that there's no \emph{best} learning algorithm \emph{only} if we average over all possible inputs to our problem.
For example, in our discussion above the label of an image $x$ is equally likely to be $y$ and $\neg y$ only if we average over all logical possibilities of such labelings. 

But such averaging is clearly silly; any label that's meaningful to a human \emph{does} correlate with the properties of the image: \Cat is a \texttt{cat}. It has whiskers and pointy ears, just like many other images of cats. Knowing the right label for this input is absolutely useful in predicting the labels of future images, since the labels are not at all random. Instead, the actual learning problems that we care about -- which are often the ones related to human intelligence -- have quite a lot of structure.

\subsubsection*{Human learning}

If we return to our exponential -- $2^n$ -- counting of the number of different $n$-pixel images,
to a human a huge fraction of these images will look like utter \emph{noise}. 
Therefore, whatever learning algorithm humans use to learn to characterize, understand, and interpret images can only ever apply to a small fraction of the possible images, and so only the tiniest fraction of the doubly-exponential potential labelings will ever be learnable by a human. Learning to classify only within that subset escapes the no-free-lunch theorem, and so humans are demonstrably able to learn things \cite{gopnik1999scientist,griffiths2006optimal,vul2014one}.

For such human learning, it has been argued that both pattern matching -- e.g.~``this image looks very similar to this other one that I've seen before, so it must also be a cat'' -- and domain specific knowledge -- e.g.~``cats have whiskers and fur and pointy ears, and so this image of an object with whiskers, fur, and pointy ears must also be a cat'' -- are essential when generalizing from sparse data \cite{tenenbaum2006theory}. Together, these make up 
the \emph{inductive bias} of a learning algorithm.
Such a bias makes \emph{a priori} assumptions about the nature of the problem, such as which features of the data are useful.

For instance, images may contain objects. Objects are made up of large components, each of which may be made up of subcomponents; recall again the relationship between whiskers and pointy ears, and the way we recognize that \Cat is a \texttt{cat}.
Images that have such a structure 
are easy to classify, both because we often understand how the components make up any object that we care about, and also because organizing objects in terms of a sequence of such \emph{representations} makes it easy for us to recognize statistical patterns among  different images at different levels of coarse-graining. 
Without this hierarchical structure, any classification task would be hopeless.

\subsubsection*{Machine learning}

Even if some of the images within the enormous set of \emph{other} images have some kind of interesting structure, a human wouldn't be able to recognize it -- in this sense, \emph{noise} is just a catch-all concept for images that are without meaning to humans. Our inductive biases only let us find \emph{human-meaningful} representations.   So how do we tell such models what gives meaning to humans?

From this perspective, the task of coming up with a learning algorithm is closely tied to the task of modeling the world. One approach suggests that we should rely on human intuition and proposes building artificial models that try to learn the way that humans learn \cite{tenenbaum2011grow,lake2017building}
rather than relying on unstructured algorithms designed to recognize statistical patterns in data. 

However, the approach based on mere ``pattern recognition'' is wildly successful \cite{ImageNet2012,goodfellow2014generative,mikolov2013efficient,mikolov2013distributed,BERT2018,radford2019language,Brown2020LanguageMA,mnih2015human,silver2016mastering,silver2018general,berner2019dota,starcraft,muzero,alphafold2prelim} despite the fact that machine learning algorithms can often find patterns in data without any human-meaningful structure. For instance, many deep learning models can memorize completely random labelings of noise \cite{zhang2016understanding}, a task that would be utterly impossible for a human.
This leaves us to wonder: what is the inductive bias of such models if they succeed at tasks that are otherwise impossible for humans, and by what mechanism are they able to %
escape the no-free-lunch theorem? 

This is why (understanding how) AI (works) is hard.

\subsection{Why is Physics simple?}\label{sec:physics-simple}

\noindent{}Physics is simple because there is no such thing as a free lunch.\footnote{Yes, that's the same lunch. Just a different interpretation. You could say that this is the unknown never-quoted statement 
that, when you average over all possible \emph{physics-learning} problems, 
all physicists are equivalent \emph{[citation needed]}.
}

That is, the reason the laws of physics are even learnable at all is because the mathematical models or \emph{theories} that offer good descriptions of the universe are especially simple models within the general frameworks that we use to enumerate the possible theories of physics. 
This is another way to interpret Wigner's observation: the laws of physics actually have an (unreasonable?) lack of \emph{algorithmic complexity}. 

Before directly addressing why our laws of physics really had to be simple in the way that they are, let's first understand where this lack of algorithmic complexity comes from.

\subsubsection*{Interacting particles}

The principal framework used to express different theories of fundamental physics is known as
\emph{quantum field theory}. %
To specify a particular theory within the framework, we first enumerate the 
degrees of freedom -- i.e.~the different elementary particles -- and then determine how they \emph{interact} with each other. 
A general interaction involves the creation of some particles and the destruction of some other particles at a particular point in space and time.

\begin{figure}[h]
\begin{center}
\includegraphics[scale=.4]{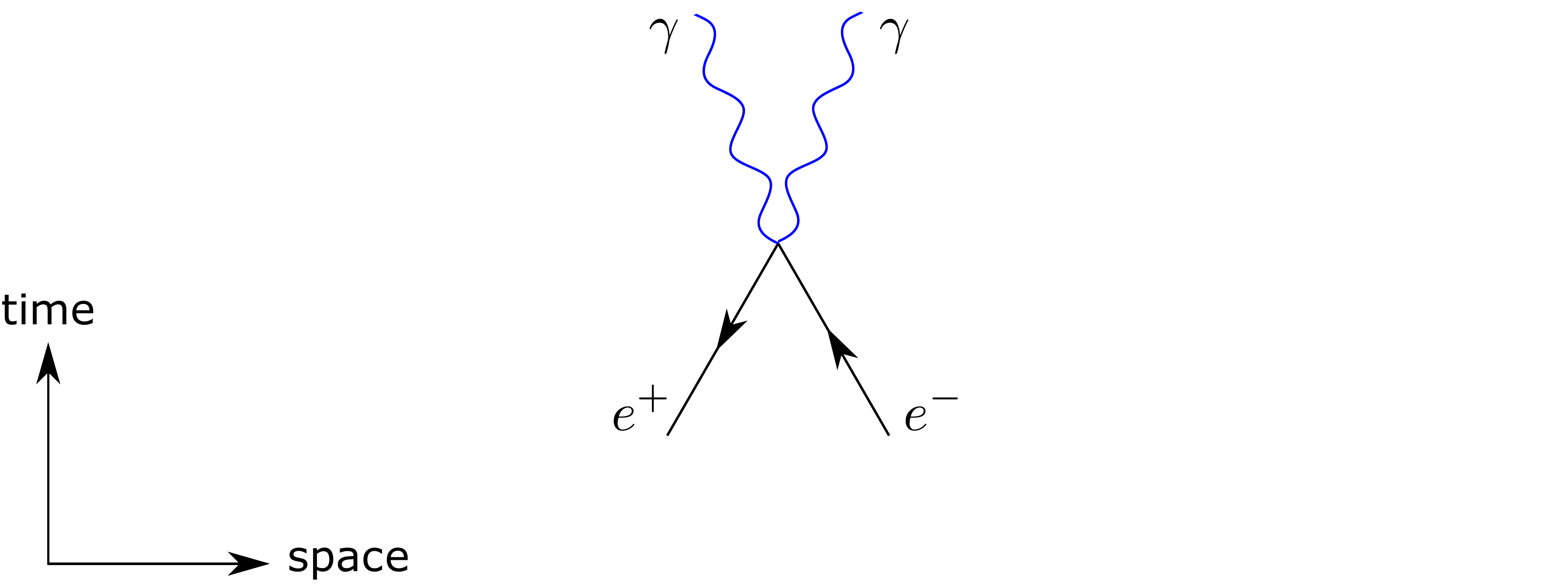}
\caption{An interaction involving $4$ particles: an electron $(e^-)$ and positron $(e^+)$ are destroyed, and 2 photons $(\gamma)$ are created.}
\label{fig:feynman}
\end{center}
\end{figure}

For example, the theory of \emph{quantum electrodynamics} is a quantum field theory that describes how matter and light interact. The basic matter particles are the electron and the positron, and the particle of light is called the photon. The strength of their interaction is set by the electric charge, which in our universe is more or less given by the dimensionless number $1/137$.  A cartoon example of such an interaction, called a Feynman diagram, 
is shown in Fig.~\ref{fig:feynman} in which an electron and positron are destroyed and two photons are created. Making use of this and many similar diagrams, an ensemble
 of grad students could predict the result of essentially any experiment involving light and matter.

The most generic theory expressed in the quantum field theory framework would have all possible combinations of interactions between all possible particles. For instance, if we have $p$ different types of particles, then
there are $\binom{p}{2}\sim p^2$ potential interactions involving $2$ particles, $\binom{p}{3}\sim p^3$ interactions involving $3$ particles, $\binom{p}{4}\sim p^4$ interactions involving $4$ particles, and so on.\footnote{We're lying here since in principle these particles can also \emph{self-interact}. (This just means that the quantum analog of the classical probability distributions that describes such particles has non-Gaussian statistics.) Taking this into account, the left-hand side of these $\sim$ expressions is false, but the right-hand side is still approximately correct.
}
Each of these interactions has a strength or \emph{coupling}, 
and a generic quantum field theory would have random nonzero values for the strength of each of these interactions.

In fact, things can get quite complicated as quantum field theories sometimes have an infinite number of degrees of freedom per particle type -- you might recognize from familiar everyday phenomena like \emph{turning up the lights} --  and a priori the strength of these interactions between all the degrees of freedom could vary in space and in time. So in some sense, the discussion in the previous paragraph was a vast under-counting of the number of potential couplings!

\subsubsection*{Typicality}

In many cases we can still effectively model such systems with a finite number $N$ degrees of freedom.\footnote{For instance, the number of states accessible at a given energy is finite, so systems with a bounded energy can be modeled with a finite number of degrees of freedom. If a concrete model is helpful for you, imagine a system of $N$ spins.}
In the absence of any guiding principle, each degree of freedom in a \emph{typical} theory can either participate or not in an interaction. Using the binomial theorem to add up all these interactions, this more or less gives $\binom{N}{2} + \binom{N}{3}  + \ldots + \binom{N}{N}\approx 2^N$ different potential interactions between the $N$ degrees of freedom. Thus, generically the number of possible interactions and therefore parameters will scale as an exponential in the size of the system.\footnote{
    If you're familiar with the Ising model, then perhaps the following is helpful: if we have $N$ spins, and there's no restriction on the interactions, then there's $2^N$ possible independent interaction terms in the Hamiltonian based on whether a particular spin is present or not in the interaction term. This is akin to a completely random Hamiltonian on $N$ degrees of freedom.
} %

As we will see, such typical theories with a number of parameters that scales exponentially with the number of degrees of freedom do not provide a good model of the physics in our universe. Concretely, the number of experiments that we'd have to perform in order to learn all those parameters is about as large as the total number of experiments we could ever perform on such a system. But if we have to perform just about every possible experiment, then such a theory has no predictive power! 

In other words, the results of one experiment would not really correlate with the results of another. Such a theory has no generalization property: learning the parameters of such a theory is tantamount to simply memorizing a look-up table with the outcomes of all such experiments.
This means that such theories are about as complicated as they could be, and the algorithmic complexity required to specify them essentially renders them useless. 

Don't worry if this reminds you a bit of the discussion in \S\ref{sec:ai-hard} -- that's typical.

\subsubsection*{Sparsity}

The key error we just made was assuming the ``absence of any guiding principle'' in determining our theory. The most important such principle is \emph{sparsity}: the number of parameters defining a theory should be far less than the number of experiments that such a theory describes. The fact that the relevant quantum field theories for the laws of physics in our universe are sparse is why they are even useful.

In the most general application of the principle of sparsity, we limit the number of degrees of freedom that can participate in any interaction. For instance, for a theory describing $N$ degrees of freedom we might limit interactions to $k$ degrees of freedom at a time, for $k \ll N$. If the cartoon in Fig.~\ref{fig:feynman} is still helpful, this would mean that we'd only need to consider diagrams with $k$ or fewer legs.

Now, let's add up the number of interactions in such a sparse theory. If the number of degrees of freedom is large ($N \gg 1$) then such a sum is dominated by the $\binom{N}{k}\sim N^k/k!$  term: the number of parameters only scales polynomially in the number of degrees of freedom.

When a sparse quantum field theory is a good model for some natural phenomenon, it's incredibly useful. If we perform the fractionally small number of experiments needed to learn all the parameters, we can then predict the result of the huge fraction of experiments we've yet to perform. From an algorithmic complexity standpoint it is somewhat miraculous that we can compress our huge look-up table of experiment/outcome into such an efficient description. 
In many senses, this type of compression is precisely what we mean when 
we say that physics enables us to understand a given phenomenon.

\subsubsection*{Locality}
The general sparsity property that we just discussed is sometimes called \emph{$k$-locality}. This emphasizes that no more than $k$ of the $N$ degrees of freedom can communicate in any given type of interaction. We impose such sparseness on our theories because it's consistent with our experimental observations.\footnote{
    The phenomenon of \emph{universality} provides an explanation for this $k$-locality, though not for spatial locality. See footnote~\ref{footnote:universality} for a brief discussion of how universality is manifested in deep learning.
}
However, it's actually far too weak a condition: typical theories that have only a sparse $k$-local structure imposed don't actually look anything like the theories that usefully describe our universe. 

What we actually observe experimentally is much stronger than $k$-locality. 
In particular, our natural intuition about the world suggests the degrees of freedom should be arranged in some manner such that some are closer together than others, and that this closeness is somehow related to their ability to interact and exchange information. This is intuitive; an experiment performed on the moon should not influence experiments performed on earth, at least not until signals from the moon containing information about such an experiment can reach the earth. 

What we need is to also impose \emph{spatial locality}. Rather than letting the $N$ degrees of freedom interact however they like so long as they're quarantined in group sizes of $k$ or fewer, we first arrange all $N$ in a lattice and then say that they can only interact with their neighbors. For instance, if we have a $1$-dimensional lattice -- a line -- each of the $N$ degrees of freedom has a left and right neighbor. More generally, if we have a $d$-dimensional hypercube then each degree of freedom will have $2d$ neighbors. To count the number of distinct interactions, we simply add up the links on the lattice and find $Nd$ total interactions: the number of possible interactions is linear in the number of degrees of freedom.

In fact, the entire concept of \emph{spatial dimension} is only meaningful because of the spatial locality of interactions! If degrees of freedom were placed at different spatial locations -- i.e.~the moon and earth -- but they could talk to each other as easily as if we placed them next to each other, then the whole notion of space itself is meaningless.\footnote{
    Relatedly, locality gives meaning to moon-landing conspiracists everywhere.
} The reason \emph{space} is a useful abstraction is because \emph{locality} is imposed on quantum field theories.\footnote{
Einstein was famously bothered by a property of quantum mechanics that naively seemed to allow actions in one place to have an effect on actions somewhere else at the same instant.
You've likely heard of this as \emph{spooky action at a distance}, though to Einstein it would have probably sounded more like \emph{spukhafte Fernwirkung} \cite{einsteinborn}.
In a paper with Podolsky and Rosen \cite{Einstein:1935rr}, Einstein was worried that this uniquely quantum phenomenon -- which is now known as \emph{entanglement} -- would violate the principle of locality. 

In particular, any local 
quantum field theory -- according to the principle of relativity -- has a rigid notion of causality that says that the outcome of an experiment at a particular point $x$ and time $t$ cannot at all influence the outcome another experiment at a different point $x'$ and time $t'$, unless the absolute distance between these experiments is less than the time interval between them times the speed of light: $|x-x'| < c|t-t'|$ \cite{Streater:1989vi}. We now appreciate that entanglement \emph{(a)} does not violate causality and \emph{(b)} arises from completely local interactions \cite{bell1964einstein}. Quantum effects are just sometimes weird and unintuitive.

Amusingly, gravitational objects known as \emph{wormholes} -- discussed by Einstein and Rosen around the same time \cite{Einstein:1935tc}  -- have a history of being confused with entanglement. A wormhole is a special type of geometry that provide a shortcut through space so that vastly separated regions can be more closely connected together. It turns out that those of us that were confused weren't actually confused, and the joke was on everyone else: both entanglement and wormholes are actually manifestations of the same underlying phenomenon \cite{Maldacena:2013xja}.
}

\subsubsection*{Translation invariance}

Our final guiding principle is called \emph{translation invariance}. Translation invariance is a \emph{symmetry} imposed on 
quantum field theories that are used to model phenomena that respect
the following empirical observation: the outcome of any experiment is generally unaffected by where it is performed. The same general laws that describe physics here on earth also describe physics in the same way on the moon and also describe physics in the same way outside our galaxy. 

To implement this principle, we require that the strength of any local interaction should be the same anywhere.\footnote{In the language of machine learning, local translationally-invariant quantum field theories enforce a very rigid notion of parameter sharing!} In other words, all the parameters have to be the same.\footnote{
    As a specific example, arranging $N$ spins into a $d$-dimensional hypercube and imposing translation invariance would leave us with the standard $d$-dimensional Ising model and only $1$ parameter.
} 
 Concretely, the electric charge we measure while at rest on the earth will be the same as the electric charge we'd measure at rest on the moon: $\approx 1/137$.

Let's spend a moment to take stock of this huge reduction in algorithmic complexity. We started with a \emph{typical} theory of $N$ degrees of freedom that requires an exponential -- i.e.~$2^{O(N)}$ -- number of parameters to specify. By imposing the most generic notion of sparsity, \emph{$k$-locality}, we reduced this to a polynomial -- i.e.~$O(N^k)$ -- number of parameters. Then by imposing $d$-dimensional \emph{spatial locality}, we reduced this to a linear -- i.e.~$O(N)$ number of parameters. Finally, by imposing \emph{translational invariance}, we could specify our theory with an $O(1)$  number of parameters. For such theories, the number of parameters no longer grows with the number of degrees of freedom: once we measure the electric charge, we can more or less predict the outcome of any other experiment involving light and matter.\footnote{That is, if we ignore the mass of the electron, which is small enough that in many cases doing so would be completely justified. Otherwise, we have an additional parameter to measure: in some sense, the electron mass can be thought of as the strength of a 2-particle ``interaction'' between an electron and a positron and \emph{no} photons.\label{footnote:electron-mass}}
The fact that the universe seems to obey these constraints is very powerful indeed!

\subsubsection*{Complexity}
To conclude our discussion of simplicity in physics, let's discuss how nontrivial complexity can arise from a sparse mathematical model.

In any sparse theory, we can study how small perturbations to some observables can grow and affect the rest of the system \cite{Lieb:1972wy,Sekino:2008he,Maldacena:2015waa}. Sparse theories have a preferred basis: the set of variables in which the formulation of the theory is sparse is special. In terms of such variables, the description of the theory is extremely simple.

In a theory with spatial locality, it's simple to perform an experiment by disturbing a local degree of freedom and observe how the system responds. The particulars of such a disturbance will grow in space and slowly spread to change the  overall state of the system: this is a quantum manifestation of chaos and the butterfly effect \cite{Larkin:1969abc,Maldacena:2015waa}.

An important property of any system -- which sometimes goes by the name \emph{thermalization} -- is how quickly it returns to equilibrium after such a disturbance. Depending on how detailed an experiment you perform, this question has a variety of different answers, ranging from a time that is independent of the number of degrees of freedom $N$ to a time that scales with the size of the system and tracks how long it takes for the disturbance to cover the entire system. If you want to find the system in a \emph{truly} random state, you might have to actually wait a time that is exponential in the number of degrees of freedom \cite{Susskind:2015toa}.

Despite this long timescale, the fine details of a random state as compared to a simple-but-perturbed state are extremely difficult to measure. In the simple-but-perturbed state, the correlations between what the original state counterfactually would have been and what the disturbed state actually is quickly grow to become more and more complicated as the difference spreads throughout the system.  However, because we can only perform local experiments, 
it's actually often difficult for the results of any such experiment to determine whether a given state is truly from a random theory or just approximately random and built up from a sparse theory \cite{Sekino:2008he,Hayden:2007cs,Lashkari:2011yi,Brandao:2012qcp,Brown:2012gy}.
That is, as far as we can actually tell, a system can be arbitrarily complicated even as it continues to change meaningfully for a long time.\footnote{For a different perspective on simplicity and complexity in physics, with connections to spin glasses and string theory, see the introduction of \cite{Denef:2011ee}.}

By contrast, without locality, there's no spatial structure, and without sparsity, there's no real structure at all. In such a completely generic theory, 
there aren't any special set of observables for experimenters to measure, track, or query.
In these systems, any disturbance of a single degree of freedom immediately grows to affect the entire system: the equilibration time is completely independent of the system size, regardless of how big it is! 
Thus, %
it's not really meaningful to observe \emph{anything}, at least in the usual sense of what we mean. In these systems, nothing happens!\footnote{The global statistical properties of these systems can have interesting time evolution see e.g. \cite{Cotler:2016fpe}. However, such observables are so overly complicated that they do not correspond to any experiments that %
could be practically performed on such systems.
} 

\subsubsection*{Physics learning}

The reason sparsity and more specifically locality are so important is because they provide structure to the space of physics theories. In fact, physical constraints due to these principles play an important role in setting which experiments can be practically carried out or algorithmically which computations are possible in the universe.\footnote{
It would be interesting to consider such constraints when assessing concerns about human existential risk due to artificial intelligence, see e.g.~\cite{bostrom2014superintelligence}.  In particular, when you ``imagine'' an agent with unbounded computational resources you should also make sure to imagine what kind of universe such an agent would have to live in. (Such a universe is entirely unlike our own \cite{Bekenstein:1980jp,tHooft:1993gx,Susskind:1994vu,Sekino:2008he,Shenker:2013pqa,Maldacena:2015waa,Brown:2015bva,Brown:2015lvg}.) For an even more speculative discussion on how physical limits to information theory may constrain or place computational limits on thinking agents, see \cite{roberts2018causality}.
}
Just as the inductive bias of human learning enables us to find \emph{human-meaningful} structure in images, sparsity and locality make up the inductive bias that lets us learn physics.\footnote{
Relatedly, since the way in which we typically interact with the world is classical -- i.e.~devoid of distinctly quantum effects -- quantum mechanics is naturally unintuitive.}  

In other words, we impose sparseness on our theories because it's consistent with our experimental observations as local observers that themselves must be described by the theory. You could say: sparseness is what the universe finds when it introspects.
Since humans are local entities within the universe, the sort of experiments that we can reason about theoretically and easily perform practically -- the sort of experiments that we have any intuition about at all -- are all based on local correlations. No wonder that the theories that we learn are designed to predict the results of local experiments!

What about all those other experiments? More or less, they are the analogues of \emph{noise}: they have no meaning to local observers like us humans. Many of them involve correlating a huge fraction of the $N \gg 1$ degrees of freedom of the system across vast regions of space. Some other experiments are impossible to perform in any reasonable amount of time by any observer within a system.\footnote{An example of such an experiment that relates to an important black hole information problem \cite{Almheiri:2012rt} is given in \cite{Harlow:2013tf}.
}  
Thus, when the observers are part of the system they are trying to learn about, the lack of algorithmic complexity of such a theory imposed by locality can privilege certain experiments and prevent others from being performed. 
The results of all such essentially impossible experiments are heavily constrained by our algorithmically simple theories of physics, though we'll never be able to check for sure!

This brings us back to the no-free-lunch theorem: the catch is that here the ``average over all possible \emph{physics-learning} problems'' is like an average over all possible experiments. And it turns out that the set of experiments that 
 actually are interesting is a tiny subset of the total number of logically possible experiments. 

 This is why physics is (able to offer) simple (explanations for seemingly 
 complicated natural phenomena).

\subsection{What can Physics do for AI?}\label{sec:do}
In summary, \emph{physics learning} -- just like \emph{human learning} more generally -- is possible because of sparsity. In the strongest but most speculative sense, such a principle is a precondition for there to be local observers capable of performing experiments. But it's still not clear from this discussion how to apply this to \emph{machine learning}.

As we mentioned, one approach to understanding machine learning is through human learning \cite{tenenbaum2011grow,lake2017building}. This is the approach of the cognitive scientist.
Such research could lead to understanding high-level principles of intelligence applicable across many models of intelligent systems, regardless of their underlying mechanisms. If the goal of machine learning is an artificial human-level intelligence, then these sorts of \emph{top-down} considerations can be incredibly useful for guiding the direction and focus of AI research.

By contrast, the physics approach is \emph{bottom up}: the focus is on understanding how the algorithmic description of a model is connected to the model's output or experimental results.
Starting from the definition of a model in terms of its microscopic degrees of freedom -- e.g.~the weights and biases of a deep neural network -- what is the high-level behavior of the model? How are these microscopic variables connected to the \emph{latent variables} in which the model has a simple description? Such an approach isn't directly concerned with the general properties of -- or constraints guiding -- human intelligence, but rather focuses on directly understanding how the specification of a model can be connected to its ability to complete a task.
In this way, the physics approach can provide a natural complement to the cognitive science approach.

Applying this bottom-up approach directly to humans is far too complicated; the path from microscopic physics through biology and neuroscience to psychology is probably too daunting for any entity short of a superintelligence\dots

Applying this bottom-up approach to machine learning, however, is likely just right. We propose that physicists interested in working out aspects of a theory of intelligence could make progress by working out a theory of current successful machine learning models through the lens of physics.
On the one hand, by studying current state-of-the-art machine learning frameworks we can study algorithms that share some properties with human intelligence.  On the other hand, the output and behavior of these algorithms aren't separated from their microscopic descriptions -- e.g.~their PyTorch \cite{paszke2017automatic} implementations -- by something as messy as biology.

The reason to be hopeful is that for these models to work at all given the no-free-lunch theorem, there must be structure. 
By trying to understand which models within a given machine learning framework -- e.g.~deep learning -- lead to sparse descriptions, perhaps we can better understand
what selects such latent descriptions from a bottom-up sparsity perspective and connect them to the top-down human perspective. %
At which point, perhaps we'll all sit at the same lunch table.

\section{The Unreasonable Effectiveness of Deep Learning}\label{sec:ETofDL}
While the discussion of physics has been varying levels of concrete, the discussion of AI has been entirely in the abstract. In an attempt to restore the balance, 
in the rest of the essay we'll focus on explaining how the framework of \emph{deep learning} can be interpreted through the lens of some of the physics principles that we've  been outlining.

While artificial intelligence and deep learning have become nearly synonymous, we haven't really discussed the specifics of deep learning at all. Deep learning is a branch of machine learning based on artificial neural networks. Such networks compute a very flexible set of functions that turn out to be excellent for mathematically modeling the sorts of tasks 
that we associate with human intelligence. 

To that point, the list of AI breakthroughs based on the deep learning framework includes problems in computer vision \cite{ImageNet2012,goodfellow2014generative}, natural language processing and generation \cite{mikolov2013efficient,mikolov2013distributed,BERT2018,radford2019language,Brown2020LanguageMA}, and the mastery of many games by deep reinforcement learning: beginning with a general purpose algorithm capable of learning to play a set of 49 Atari games \cite{mnih2015human}, and growing to include Go \cite{silver2016mastering,silver2018general}, chess without a specialized algorithm \cite{silver2018general}, Dota II \cite{berner2019dota}, Starcraft \cite{starcraft}, and even a unifying algorithm for handling Go, chess, shogi, and 57 Atari games without any prior knowledge of the dynamics of those games \cite{muzero}. 
Most recent -- and perhaps most exciting  -- is a deep learning solution to the protein folding problem in biology \cite{alphafold2prelim}.

However, deep learning's effectiveness is puzzling.
In defiance of the cognitive science perspective, domain specific knowledge has often proved neutral if not harmful in guiding the approach of practitioners. While the choice of the specific model -- the neural network architecture -- 
is often essential, within a class of architectures that work the most important considerations leading to success always seem to be the size of the model -- huge -- and the amount of training data -- as much as possible \cite{sutton_2019}.\footnote{For instance, in the domain of natural language modeling the advent of the transformer architecture \cite{attention2017} has arguably rendered much of the expertise of natural language processing researchers obsolete. Instead, the most important ingredient of success is ever larger models and ever more data \cite{kaplan2020scaling}.
}
Moreover, unlike human learning models, deep learning models that have sufficiently many parameters can also memorize random labelings of random data \cite{zhang2016understanding}.
In short, attempts to apply top-down analysis to explain the success of deep learning often come up short.

This suggests there might be something to gain by a physicist's bottom-up analysis. 
In \S\ref{sec:introduction}, we argued that \emph{sparsity} is an essential guiding principle of physics theory. With that discussion in mind, in this section we will similarly argue that sparsity is a guiding principle of deep learning theory.

The material presented here is an overview of the perspective taken in an upcoming book with Yaida and Hanin \cite{Principles}.

\subsubsection*{Interacting neurons}\label{sec:EFT}

\epigraph{On being asked, ``How is Perceptron performing today?'' I am often tempted to respond, ``Very well, thank you, and how are Neutron and Electron behaving?''}{Frank Rosenblatt, inventor of the perceptron and also the Perceptron  \cite{rosenblatt1961principles}.}

\noindent{}A very natural framework for studying deep learning is closely related to the sort of quantum field theory that we described in \S\ref{sec:physics-simple}, with the first principal difference that deep learning is a classical statistical theory, not a quantum theory, and the second principal difference that the statistical variables in deep learning are not organized into fields. 

At the highest level, this doesn't matter. For both deep learning and quantum field theory there is an organizing principle called \emph{effective theory} that allows us to make progress.

To begin, we need to specify the degrees of freedom in deep learning, just as we did before when discussing quantum field theory. Rather than Electrons and Neutrons (or rather, electrons and neutrons), in this case we have neurons.\footnote{
    The perceptron model, invented by Rosenblatt, was the first artificial neural network with learnable weights \cite{rosenblatt1958perceptron}.
} 

Neurons take a weighted sum of input signals and then fire if that sum is above a certain threshold.\footnote{
    More generally, the weighted sum of input signals minus the threshold is acted on by a scalar \emph{activation function}. For the \emph{perceptron} activation function \cite{mcculloch1943logical}, this corresponds to the simplistic notion of firing.
} In a typical neural network, many such neurons are organized into layers, and deep learning places a particular emphasis on networks that are iteratively composed of many layers with a similar structure. 

\begin{figure}[h]
\begin{center}
\includegraphics[scale=.9]{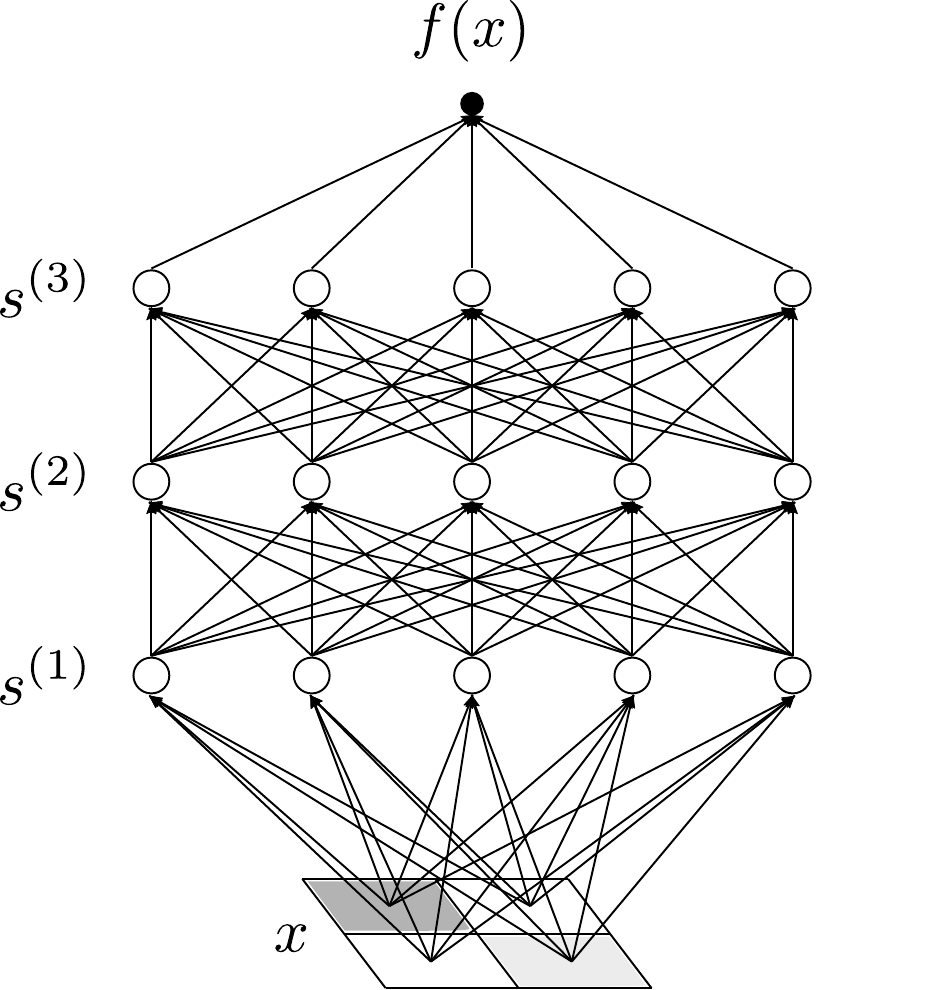}
\caption{Depiction of a simple multilayer neural network. The circles represent the neurons in the hidden layers, the black dot at the top represents the network output, and the lines indicate the connections between the neurons. This four-layer network takes a $2 \times 2$ image $x$ and computes the function $f(x)$ after passing the input through 3 hidden layers. The input is transformed into the output at layer $L=4$ through a sequence of intermediate signals, $s^{(1)}$, $s^{(2)}$, and $s^{(3)}$.
}
\label{fig:mlp}
\end{center}
\end{figure}

An example of such a multilayer structure is shown in Fig.~\ref{fig:mlp}. In the network represented by this figure, the function $f(x)$ is computed by sequentially passing signals through the layers of neurons. The input $x$ is transformed into the signal $s^{(1)}$ by the five neurons in the first hidden layer, then into a sequence of signals $s^{(2)}$ and $s^{(3)}$ by the second and third hidden layers, respectively, before being transformed into the output $f(x) \equiv s^{(4)}$ at the single-neuron final layer (black dot at the top). Each neuron $i$ in a layer $\ell$ forms a component of the signal in that layer, e.g.~$s_i^{(\ell)}$. Moreover, an important quantity of interest is the number of layers or \emph{depth} of the network.

The strength of the connections between the layers can be individually modified by a group of parameters called \emph{weights}. The firing threshold of each neuron is set by another set of parameters called \emph{biases}. Together, the weights and biases make up the model parameters.
In a typical learning algorithm,  these parameters are iteratively adjusted until the network output approximates some desired function on a sequence of training examples $\mathcal{D} = \{x_1, \dots, x_{N_\mathcal{D}} \}$. 

On the one hand, after such a \emph{training} procedure, the graph gives a simple way to compute the function $f(x)$ given the settings of the weights and biases. On the other hand, knowing the network architecture and the trained values of the model parameters doesn't really offer any insight into how the function works. In terms of these 
variables, the network behaves like a \emph{black box}.

Consider the language model GPT-3, which has 175 billion parameters \cite{Brown2020LanguageMA}. Assuming the values of these parameters are stored as four-byte floats, that's 700 gigabytes! If someone hands you those 700 GB and asks you to predict the output of the model on a given input $x$, the best thing to do would be to 
explicitly compute the function $f(x)$ using an implementation of the model. If someone asks you \emph{why} the output $f(x)$ is given on an input $x$, you're in trouble; these microscopic variables -- the model parameters -- are just not the right set of variables in which the computation is easy to understand.\footnote{This is like being given a description of a person's \emph{connectome} and asked to predict what such a person would say if prompted with the input $x$; the best you can hope to do is actually go find the person and ask them $x$!}

However, there's another way of thinking about what the network is doing.\footnote{
    More precisely, by sampling the weights and biases from an initialization distribution, we are \emph{inducing} a distribution on the network outputs as well as the on the hidden layer signals. Then, to analyze this \emph{ensemble}, we formally integrate out the weights and biases. 
} For the output neuron defining the function, there's some marginal probability that it fires on a particular input: $p(f | x)$. 
More generally, for the training dataset $\mathcal{D} = \{x_1, \dots, x_{N_\mathcal{D}} \}$, there's a joint probability distribution encoding the probabilities that the output fires on a selection of inputs: %
\be\label{eq:network-output}
p(f | \mathcal{D}) \equiv p\le(f(x_1), \dots, f(x_{N_\mathcal{D}}) \ri) \, . 
\ee
This \emph{output distribution} encodes the correlations of the network output when the network is evaluated on different inputs.
Even more generally, we could consider a sequence of $L$ \emph{neural distributions} 
\be\label{eq:neural-distribution-sequence}
p(s^{(1)} | \mathcal{D}),~ p(s^{(2)} | \mathcal{D}),~   \dots,~ p(s^{(L)} | \mathcal{D})\, ,
\ee 
characterizing how the input data $\mathcal{D}$ is transformed into the network output $f(x) \equiv s^{(L)}$ at layer $L$ through the signals $s^{(\ell)}$ computed by the $\ell$-th layer neurons in each of the hidden layers.

In a way that can be made quite precise \cite{Principles}, we can interpret these correlations in these neural distributions as arising from \emph{interactions} between the neurons just as we described quantum field theory in terms of the interactions of elementary particles. Working out the form of each of the distributions in this sequence lets us understand how a network transforms input into output; it is tantamount to opening the black box of deep learning. So, as you can imagine, it's naively a pretty difficult thing to do.

Mirroring our discussion of the quantum field theory degrees of freedom before, the most generic neural network expressed in the deep learning framework would have all possible interactions between all the different neurons. Focusing  on just the network output for a moment, if we have $N_\mathcal{D}$ different inputs in our dataset,  then there are $\binom{N_{\mathcal{D}}}{2}\sim N_\mathcal{D}^2$ potential interactions involving the output neuron evaluated on two inputs, $\binom{N_{\mathcal{D}}}{3}\sim N_\mathcal{D}^3$ potential interactions involving the output neuron evaluated on three different inputs, $\binom{N_{\mathcal{D}}}{4}\sim N_\mathcal{D}^4$ potential interactions on four different inputs, and so on. Each of these interactions has a strength or \emph{data-dependent coupling} which encodes the patterns of correlations in the input data at the network output after being transformed by the neural network.\footnote{
    These couplings should not be confused with the \emph{parameters} -- the weights and biases -- of the network. Instead, the data-dependent couplings give a alternate description of the underlying model.
}

Adding up all these interactions, this more or less gives $2^{O(N_\mathcal{D})}$ different neural interactions, meaning that a generic description of the output distribution of the network in terms of these interactions will scale exponentially with the size of the dataset. Such a description is certainly not sparse: depending on how the total number of weights and biases compares to $2^{O(N_\mathcal{D})}$, this is extremely likely to be even more complicated than the microscopic description of the network in terms of the model parameters!

\subsubsection*{Sparsity in the thermodynamic limit} %

One of the most powerful tools in theoretical physics is a certain asymptotic limit where the number of degrees of freedom of a system is taken to be very large. 
This idea has many different manifestations and as a result carries a number of different names: the thermodynamic limit, the semi-classical limit, and large-$N$. %

Given the many different avatars, there are many different ways to think about such a limit, but the central idea is that when the number of degrees of freedom of a system are formally taken to be infinite, the system's interactions can turn off 
and 
as a consequence
\emph{fluctuations} of the degrees of freedom are heavily suppressed.\footnote{
    Among the many systems for which this limit leads to a simple description, a familiar one is the thermodynamic description of gas in a room in terms of the ideal gas law.
}
This is essentially a physics instantiation of the central limit theorem.

In the thermodynamic limit complicated interacting systems often exhibit extremely simple behavior, and deep learning is no exception. In 1996, Neal described such a limit for neural networks with only a single hidden layer by taking the number of neurons in the hidden layer to be infinite \cite{neal1996priors}. 
This same limit was more or less employed in the multilayer setting by numerous authors \cite{poole2016exponential,raghu2017expressive, schoenholz2016deep,lee2018deep,g2018gaussian,jacot2018neural}. In this case, the number of neurons $N_\ell$ in each hidden layer $\ell$ is taken to be infinite, while the overall depth of the network $L$ is held fixed.

This \emph{infinite-width limit} provides a simple and tractable \emph{toy model} of deep learning. In particular, in this limit all the neural interactions turn off. 
This means that  the sequence of neural distributions \eqref{eq:neural-distribution-sequence} -- and in particular the output distribution \eqref{eq:network-output} -- will each converge to multivariate Gaussian distributions.
For each of these distributions, the pattern of correlation is determined entirely by the distribution's pairwise covariance matrix.\footnote{
    In the language of footnote~\ref{footnote:electron-mass}, we can alternatively think of this covariance as defining a $2$-neuron interaction just as we thought of the electron mass as a $2$-particle interaction between an electron and positron. The point is that such a $2$-particle correlation is exceptionally simple and allows the degrees of freedom to become disentangled: there always exists a basis in which their \emph{statistical independence} is manifest.
}

Thus, the infinite-width limit leads to exceptionally sparse representations: we've reduced the number of data-dependent couplings required to describe the network's output distribution from scaling exponentially in the size of the dataset $\sim2^{O(N_\mathcal{D})}$ to scaling quadratically $\sim O(N_\mathcal{D}^2)$. Naively, while such models seem \emph{overparameterized} -- potentially containing more parameters $N \to \infty$ than training data $N_\mathcal{D}$ -- %
in terms of the data-dependent couplings, they are actually sparse!

Further theoretical analysis, however, shows that this limit is too simple: these networks only permit a very simplified notion of learning in which the features used to determine the network output are fixed before any training begins \cite{jacot2018neural,chizat2018note, li2019towards}. Instead, only the coefficients of a linear function of those fixed random features get modified during training, severely constraining the classes of functions that can be learned.
To understand why this is problematic, let us recall our discussion of \emph{human learning} in \S\ref{sec:ai-hard}. There, we argued that understanding data in terms of a sequence of 
representations was an essential component of human learning;
a similar mechanism is supposed to be an essential component of \emph{deep learning} as well \cite{lecun2015deep,Rob}.

In the typical discussion of representation learning, we start with the fine-grained representation of an input such as an image in terms of its pixels:  $x =\Cat$. For a classification task, a network might output a coarse-grained representation of that image: $f(x) = \texttt{cat}$. In between, the signals at the hidden-layer neurons $s^{(\ell)}$ form intermediate representations. For instance, the initial layers can act as oriented edge detectors, while the deeper layers form more and more coarse-grained representations, organizing human-meaningful sub-features such as \texttt{fur} and \texttt{whiskers} into higher-level features like a \texttt{face}.

However, the intermediate representations in infinite-width networks are fixed from the start, completely independent of the training data. 
In a sense, the behavior of networks in this infinite-width limit is very similar to the behavior of networks without any hidden layers. By virtue of being shallow, such networks don't contain intermediate representations of their input data, and their output is always described by a Gaussian distribution. In other words, in the infinite-width limit networks are neither \emph{deep} nor do they \emph{learn} representations.  

The lack of \emph{representation learning} in the infinite-width limit indicates the breakdown of its usefulness as a toy model. %
This breakdown thus hints at the need to go beyond such a limit in order 
to describe deep networks at any nontrivial depth.

\subsubsection*{Sparsity and complexity in the \texorpdfstring{$1/N$}{1/N} expansion}%
\epigraph{Everything should be as simple as it can be, but not simpler.}{Roger Sessions, simplifying a much longer quote of Albert Einstein  \cite{sessions_1950}.}

\noindent{}Toy models -- like the infinite-width network -- are invented to strip away irrelevant details in order to more easily understand the essential properties of an underlying phenomenon.\footnote{
All mathematical models are toy models or approximations, in one way or another. For example, consider the Heliocentric model of the solar system with its fixed sun and non-interacting planets, or consider the Ising model as a simplistic theoretical description of ferromagnetism. The Standard Model of particle physics \cite{Glashow:1961tr,Weinberg:1967tq,Salam:1968rm} -- perhaps the most successful mathematical theory of natural phenomena in terms of the detailed correspondence of its theoretical predictions to the outcomes of experiments -- is just a very rich toy model: 
it does not include the force of gravity.
Typically, the adjective ``toy'' is reserved for overly-simplistic models.
}
When a toy model breaks down, it indicates the correspondence between the mathematical model and the underlying physical reality diverges.

For this reason, it's sometimes useful to consider a sequence of toy models with increasing sophistication.\footnote{
    For instance, Newtonian gravity, general relativity, and string theory are a sequence of increasingly sophisticated models of gravity.
} 
The probing of a rigid framework 
    with a sequence of such models leads to a built-in type of \emph{transfer learning} between the models; it's often clear from the start -- or after a quick calculation -- how a particular model can be related to others in the sequence.
 Moreover, we can often
concretely connect the 
simplifying properties used to construct these models
to the situations in which they break down.  
This is tantamount to actually understanding the role that such properties play across a broad class of models living in the underlying framework.

The program of constructing such a sequence of models in the framework of deep learning was first begun by Yaida in \cite{Yaida2019}. To define these models, we will make use
of an ancient trick in physics for studying small deviations from idealized behavior: 
\emph{perturbation theory}. 
Traditionally, perturbation theory applies to systems that have an infinitesimally-small dimensionless parameter $(\epsilon \ll 1)$ in terms of which any quantity of interest can be Taylor expanded: $f(\epsilon) \approx f(0) + \epsilon f'(0) +  \epsilon^2 f''(0)/2 + \dots$.
In this case we will also perform a magic trick -- for it is really something of a sleight-of-hand -- by realizing that perturbation theory can also be applied to systems that have an infinitely-large dimensionless parameter $(N \gg 1)$ by defining $\epsilon \equiv 1/N$. This is sometimes called the \emph{$1/N$ expansion} \cite{tHooft:1973jz}.

In particular, the $1/N$ expansion lets us back off of the infinite-width thermodynamic limit and compute \emph{finite-width corrections} to the network output distribution \eqref{eq:network-output} for networks that have a finite number of neurons per layer. Concretely, let's imagine the number of the neurons $N_\ell$ in each of the hidden layers $\ell = 1, \dots, L-1$ are all equal: $N_\ell \equiv N $. Then, we can compute the network output distribution by working out a perturbation series in the small parameter $1/N$, expanding around the $N\to \infty$ infinite-width toy model. Each toy model in the sequence is then defined according to the order in $1/N$ at which the perturbation series is truncated. By retaining additional higher-order corrections, we get richer toy models, asymptotically approaching a complete description of the underlying network.\footnote{
    The sequence of toy models in the $1/N$ expansion should not be confused with the sequence of neural distributions \eqref{eq:neural-distribution-sequence} computed by the layers of a network. In particular, each toy model gives an explicit representation of all the distributions \eqref{eq:neural-distribution-sequence} truncated to a fixed power of $1/N$.
}

Importantly, the finite-width corrected output distributions are no longer multivariate Gaussian distributions. Each order in $1/N$ turns on an additional set of neural interactions: the $O(1/N)$ correction allows interactions between collections of four neurons, each evaluated on potentially different inputs; the $O(1/N^2)$ correction allows interactions between collections of six neurons, each evaluated on potentially different inputs; and generally the $O(1/N^k)$ correction would allow interactions between collections of $(2k+2)$ neurons.\footnote{
    For a technical reason, there are only interactions between even numbers of neurons.
}

Accordingly, these distributions are defined in terms of more and more intricate patterns of correlation. At leading order in $1/N$, describing the four-neuron interaction requires $\sim N_{\mathcal{D}}^4$ data-dependent couplings, and generally describing a $(2k+2)$-neuron interaction requires $\sim N_{\mathcal{D}}^k$ data-dependent couplings.
Thus, these models populate the entire spectrum from sparse to complex! 

Despite the increasing complexity, these models also have \emph{less} naive parameters -- less weights and less biases -- than their infinite-width counterparts. Taking $N$ to be finite reduces the number of neurons per layer -- from infinite to finite -- but makes the effective description -- in terms of the data-dependent couplings encoding the pattern of neuron-neuron correlations -- more complicated and less sparse! %

Going forward our focus will be on the simplest \emph{nearly-Gaussian} model at $O(1/N)$,  including only the leading finite-width corrections. Such a model is still quite sparse -- requiring $\sim N_{\mathcal{D}}^4$ data-dependent couplings -- but also is potentially  rich enough to capture representation learning in deep networks.

\subsubsection*{Representation learning from renormalization group flow}\label{sec:rg}

To investigate representation learning in our finite-width model, we need to introduce one final tool from theoretical physics: \emph{renormalization group flow} \cite{PhysRevB.4.3174,PhysRevB.4.3184}. The Wilsonian renormalization group (RG) flow equations detail how the strength of interactions can change with the scale at which an experiment is carried out. Recall from our discussion of quantum electrodynamics in \S\ref{sec:physics-simple} that we kept specifying the strength of the electric charge, $\approx 1/137$,  \emph{at rest}. More precisely, this coupling can change or \emph{run} depending on the distance or energy scale at which matter and light are interacting.

The \emph{flow} in \emph{RG flow} is generated by a repeated marginalization over the microscopic fine-grained degrees of freedom of the system, which results in an \emph{effective theory} of coarse-grained observables.
In the language of RG flow, interactions that grow with the flow are called \emph{relevant}, and interactions that decrease in strength are called \emph{irrelevant}.
By focusing only on the growing relevant interactions, the effective theory provides an efficient way to study the dynamics of the course-grained degrees of freedom.

In the framework of deep learning, we can derive analogous flow equations -- in this case discrete recursions -- that detail how the neural interactions run as we change the depth of the network from $L$ layers to $(L+1)$ layers \cite{Yaida2019}. Moreover, these recursions tell us how to relate the sequence of neural distributions \eqref{eq:neural-distribution-sequence} describing the hidden-layer representations. Such recursions constrain the data-dependent couplings of these distributions and thus describe how the pattern of correlation in the data evolves from the input through the hidden-layer representations to the network output.

From this perspective, it's easy to make the connection between these recursions and RG flow concrete. To compute the layer $(\ell+1)$-th distribution $p(s^{(\ell+1)}|\mathcal{D})$, in the sequence of neural distributions \eqref{eq:neural-distribution-sequence}, we first construct the interlayer joint neural distribution between neurons in the $(\ell+1)$-th layer and the $\ell$-th layer,  
$p(s^{(\ell+1)}, s^{(\ell)}|\mathcal{D})$, and then marginalize over the neurons in the $\ell$-th layer $s^{(\ell)}$. Repeating such a marginalization induces an RG flow, integrating out the fine-grained representations of the shallower layers in favor of the coarse-grained representations in the deeper layers. 

In the context of deep learning, we call this \emph{representation group flow}, or \emph{RG flow} for short \cite{Principles}. Such an RG flow is precisely what we were interested in from the onset: the flow lets us understand how the microscopic variables, such as the pixels of the input image, are transformed into intermediate representations in the hidden layers, and then how those hidden-layer representations determine the network output.

With enough effort, the discrete RG flow equations can be worked out for any of the toy models described by the $1/N$ expansion. With orthogonal effort, such equations can be analyzed. In particular, by considering the asymptotic limit of large network depth $L$, we can understand the behavior of networks that are both deep and wide.

Solving the discrete RG flow equations at large $L$, we find that the finite-width corrections are \emph{relevant}, growing increasingly larger as the depth of the network increases. Explicitly, we find that leading $1/N$ corrections also scale linearly with $L$. This suggests that the proper perturbative parameter defining our sequence of toy models actually should have been the depth-to-width ratio of the network: $L/N$. This aspect ratio $L/N$ behaves as an \emph{emergent scale} or \emph{cutoff}, controlling the importance of the finite-width corrections to the infinite-width limit.

Moreover, this analysis lets us understand what went wrong in the infinite-width limit. Recall in that limit that we effectively took the number of neurons $N$ in each hidden layer to be infinite while holding the overall depth of the network $L$ fixed. This means that the ratio vanishes: $L/N \to 0$. Formally, in this limit networks behave as if they are not at all deep. Accordingly, our naive notion of what it means for a realistic network to be wide $N \gg 1$ was wrong; instead we should have been comparing width to depth.
Now we see that clearly that the proper way to study corrections to the infinite-width limit is large depth $L$ and large width $N$, with their ratio $L/N$ held fixed.\footnote{
    A very similar thing happens for the \emph{strong interactions} in physics. In the limit of Yang-Mills theory as the number of colors goes to infinity $(N\to\infty)$, you get a very different limit depending on how the Yang-Mills coupling $g$ scales with $N$ \cite{tHooft:1973jz}. Depending on this scaling, you can get a trivial free theory -- analogous to the infinite-width limit -- or a theory where the interactions are too strong to admit a $1/N$ expansion. Naturally, with the correct scaling you find string theory \cite{Maldacena:1997re}. 
}

This RG flow analysis also underscores the importance of paying attention anytime a dimensionless scale grows infinitely large or becomes perturbatively small. Further analysis of the details of initialization and training shows that scales such as width $N$ and depth $L$ are typically responsible for the extreme range of optimal hyperparameter tunings found by deep learning practitioners \cite{Principles}. The varying of such parameters over many orders of magnitude signals to an effective theorist that the cutoff of a theory is not properly understood. In particular, most hyperparameters -- appropriately constructed -- should be order-one quantities, robust to small changes and applicable across a variety of different architecture widths and depths.

Of course, the lack of representation learning in the infinite-width limit was also indicative of the breakdown of the infinite-width limit. By carefully studying the leading $L/N$ finite-width corrections, we can incorporate the effects of depth at finite width. In this limit, not only do we see nontrivial representation learning, we find that it's \emph{relevant} \cite{Principles}. In other words, we explicitly can see how \emph{deepness} leads to representation \emph{learning}!

\subsubsection*{Simplicity from criticality}

So why is such a simple toy model -- in particular, the leading finite-width nearly-Gaussian model -- so effective at describing deep learning? Are the other richer models that incorporate higher-order corrections from the $1/N$ expansion capturing other qualitative features of deep learning that the simplest finite-width model is missing? 

We can make progress on these questions by appealing to the physics principle of \emph{criticality}. First, note that for iterative maps like the one that defines the neural network shown in Fig.~\ref{fig:mlp}, exponential growth or decay of signals is generic. Such behavior is actually quite harmful to the performance of the network: exponential decay means that input signals effectively disappear after some characteristic depth scale, while exponential growth leads to numerical instability.\footnote{
    Concretely, the network targets are typically $O(1)$ numbers, so exponential behavior makes training exponentially difficult \cite{poole2016exponential,raghu2017expressive,schoenholz2016deep}.
}

However, systems tuned to a \emph{critical point} exhibit self-similar behavior under RG flow. Thus, the principle of criticality suggests that we should search for nontrivial fixed points of the RG flow equations in which the relevant observables do not behave exponentially.  In \cite{Principles} we explain how to tune the initialization distributions of the weights and biases of the network order by order in the $1/N$ expansion in order to reach criticality.\footnote{
One consequence of renormalization group flows to nontrivial fixed points is the phenomenon of \emph{universality} \cite{Kadanoff:1971pc}. The idea is that the sequence of distributions describing different systems at various levels of coarse-graining can converge under RG flow, despite having potentially very different fine-grained microscopic descriptions. The set of such systems that behave similarly are said to form a \emph{universality class}.

Similarly for \emph{representation group flows}, we find that the sequence of neural distributions \eqref{eq:neural-distribution-sequence} can also converge under RG flow for networks employing different activation functions \cite{Principles}. For activation functions within a universality class, the details 
become irrelevant, with the behavior of observables explicitly depending on only a few Taylor coefficients of the function. \label{footnote:universality}
}

With that in mind, one feature of critical systems is the manifestation of exceptionally large \emph{fluctuations} on the order of the size of the system.\footnote{
     Recalling our discussion of the thermodynamic limit, it's clear that there's some tension between this notion of criticality and the infinite-width limit in which  fluctuations are supposed to be suppressed. This offers another way of thinking about the infinite-width limit: it's generically an unstable fixed point of the RG flow.
}  In the context of deep learning, this means large instantiation-to-instantiation fluctuations between explicit initializations of a network's weights and biases. The RG flow analysis for networks at criticality shows that these fluctuations have a typical size that scales with the emergent cutoff $\sim L/N$. When such fluctuations are large, destructive exponential behavior is generic in any particular instantiation.

Thus, on the one hand, to suppress such fluctuations we require $L/N < 1$. On the other hand, to get nontrivial representation learning, we require $L/N$ nonzero. That is, successful networks are both wide and deep, with their aspect ratio reasonably small but nonzero. In this regime, the simplest finite-width-corrected \emph{nearly-Gaussian} model thus captures the essence of deep learning, realizing the paraphrased Einstein's principle of simple, but not too simple.

\subsubsection*{Deep learning: an effective theory approach}

To review, our goal was to find a physicist's bottom-up formalism for understanding the principles of deep learning theory. In the infinite-width limit, we observed that all the neural interactions turned off, leading to a very sparse toy model of neural networks with limited use. This further let us explain how \emph{overparameterized} networks with more tunable parameters than training data are secretly really simple in terms of their sparse \emph{data-dependent coupling} descriptions.

Backing off the strict infinite-width limit, we found a nontrivial expansion in the depth-to-width ratio $L/N$ that lets us systematically incorporate interactions into the theory at the cost of a more complicated analysis. Non-intuitively, reducing the naive number of parameters -- by taking finite $N$ %
 -- is actually a means of making our theoretical description \emph{less sparse}!

Then, the RG flow analysis of the $1/N$ expansion gave an effective theory understanding of representation learning, while criticality offered insight into the tuning of hyperparameters. All together, this makes our effective theory approach \cite{Principles} extremely useful for understanding real neural networks in practice.

\section{The Future}\label{sec:future}

\epigraph{TANSTAAFL! -- \emph{Robert A. Heinlein \cite{heinlein}.}}{}\vspace{-1\baselineskip}

\noindent{}There might not be free lunches, but there will be lunch specials.

\section*{Acknowledgments}
We are grateful to Yasaman Bahri, John Frank, Yoni Kahn, Yann LeCun, Kyle Mahowald, George Musser, Adrienne Rothschilds, David Schwab, Douglas Stanford, DJ Strouse, Josh Tenenbaum, and Jesse Thaler for discussions and feedback. We are especially grateful to Boris Hanin and Sho Yaida for collaboration, discussions, and for providing extensive feedback on multiple drafts.
This essay was brought to you in the limit of $L/N$ held fixed, and 
by readers like you.
Thank you for your time.

\mciteSetMidEndSepPunct{}{\ifmciteBstWouldAddEndPunct.\else\fi}{\relax}
\bibliographystyle{utphys}
\bibliography{why}{}

\end{document}